\documentclass[particles,review,accept,pdftex,moreauthors]{Definitions/mdpi}

\firstpage{417}
\pubvolume{7}
\issuenum{2}
\articlenumber{24}
\pubyear{2024}
\copyrightyear{2024}
\externaleditor{Academic Editor: Armen Sedrakian}
\datereceived{20 December 2023}
\daterevised{29 March 2024} % Comment out if no revised date
\dateaccepted{11 April 2024}
\datepublished{20 April 2024}
\hreflink{https://doi.org/\linebreak 10.3390/particles7020024} % If needed use \linebreak
\pdfoutput=1 % Uncommented for upload to arXiv.org
\usepackage[utf8]{inputenc}
\usepackage{epsfig}
\usepackage{setspace}
\usepackage{setspace}
\usepackage{epsfig}
\usepackage{url}
\usepackage{multirow}
\usepackage{array}
\usepackage{tabularx,colortbl}
\usepackage{longtable}
\usepackage{lscape}
\usepackage{pdflscape}
\usepackage{changepage}
\usepackage{color}
\usepackage{ragged2e}
\usepackage{booktabs}

\definecolor{red}{rgb}{1,0,0}
\definecolor{green}{rgb}{0,0.5,1}
\definecolor{cinza}{rgb}{0.5,0.5,0.5}

\Title{Feature Selection Techniques for CR Isotope Identification with the AMS-02 Experiment in Space}

\TitleCitation{Feature Selection Techniques for CR Isotope Identification with the AMS-02 Experiment in Space}

\Author{Marta 
Borchiellini $^{1,}$*\href{https://orcid.org/0009-0006-3805-2983}{\orcidicon}, Leandro Mano $^{2}$\href{https://orcid.org/0000-0003-2215-0133}{\orcidicon}, Fernando Bar\~ao $^{3,4}$\href{https://orcid.org/0000-0002-8346-9941}{\orcidicon} and Manuela Vecchi $^{1}$\href{https://orcid.org/0000-0002-5338-6029}{\orcidicon}}

\AuthorNames{Marta Borchiellini, Leandro Mano, Fernando Bar\~ao and Manuela Vecchi}

\AuthorCitation{Borchiellini, M.; %MDPI: Please check all author names carefully.
Mano, L.; Bar\~ao, F.; Vecchi, M.}
\address{%
$^{1}$ \quad Kapteyn Astronomical Institute, University of Groningen Landleven 12,  9747~AD~Groningen,~The~Netherlands; m.vecchi@rug.nl\\
$^{2}$ \quad National Council of Scientific and Technological Development, SHIS Q1, Edifício Santos Dumont 203, Brasília~71605-001,~Brazil; %MDPI: Please provide the name of the city and postal code (or ZIP code in the U.S.). If the postal code is not available, Post Office Box number can be added instead.
leandroyukiomano@gmail.com\\
$^{3}$ \quad Laboratório de Instrumentação e Física Experimental de Partículas (LIP), 1649-003 Lisboa, Portugal; barao@lip.pt \\
$^{4}$ \quad Departamento de F\'\i sica, Instituto Superior T\'ecnico---IST, Universidade de Lisboa---UL, Avenida Rovisco Pais~1, 1049-001 Lisboa, Portugal}

\corres{Correspondence: m.borchiellini@rug.nl} %Tel.: (optional; include country code; if there are multiple corresponding authors, add author initials) +xx-xxxx-xxx-xxxx (F.L.)

\abstract{Isotopic composition measurements of singly charged cosmic rays (CR) provide essential insights into CR transport in the Galaxy. The Alpha Magnetic Spectrometer (AMS-02) can identify singly charged isotopes up to about 10 GeV/n. However, their identification presents challenges due to the small abundance of CR deuterons compared to the proton background. In particular, a high accuracy for the velocity measured by a ring-imaging Cherenkov detector (RICH) is needed to achieve a good isotopic mass separation over a wide range of energies.
The velocity measurement with the RICH is particularly challenging for $Z=1$ isotopes due to the low number of photons produced in the Cherenkov rings. This faint signal is easily disrupted by noisy hits leading to a misreconstruction of the particles' ring. Hence, an efficient background reduction process is needed to ensure the quality of the reconstructed Cherenkov rings and provide a correct measurement of the particles' velocity.
Machine learning methods, particularly boosted decision trees, are well suited for this task, but their performance relies on the choice of the features needed for their training phase. While physics-driven feature selection methods based on the knowledge of the detector are often used, machine learning algorithms for automated feature selection can provide a helpful alternative that optimises the classification method's performance.
We compare five algorithms for selecting the feature samples for RICH background reduction, achieving the best results with the Random Forest method. We also test its performance against the physics-driven selection method, obtaining better~results.}

\keyword{feature selection; cosmic rays; isotope identification}%(List three to ten pertinent keywords specific to the article; yet reasonably common within the subject discipline.)

\begin{document}

%\begin{spacing}{1.5}

% Citing a journal paper~\cite{ref-journal}. Now citing a book reference~\cite{ref-book1,ref-book2} or other reference types~\cite{ref-unpublish,ref-communication,ref-proceeding}.
%All figures and tables should be cited in the main text as Figure~\ref{fig1}, Table~\ref{tab1}, etc.

\section{Introduction}
\label{sec:1}

Positive, singly charged nuclei dominate the galactic cosmic ray (CR) spectrum~\cite{gaisser}. Cosmic rays can be divided into two main categories based on their production mechanism: primary cosmic rays, which are produced directly in stellar nucleosynthesis processes at the sources, and secondary cosmic rays, which originate from the nuclear interaction of primary CRs with the interstellar medium (ISM) during their propagation in the Galaxy~\cite{coste}. Although protons dominate the isotopic composition of hydrogen in cosmic rays, a few percent of deuterons are also present. They are expected to be mostly of secondary origin since the primary deuterons produced in the first step of the proton--proton chain are depleted in the next step of the nucleosynthesis reaction~\cite{ppchain}. Secondary deuterons are produced through inelastic interactions between CRs, mainly p, $^3$He, $^4$He, and the ISM. Thus, the identification of deuterons and the measure of their flux is essential for the study of cosmic ray propagation processes in the Galaxy. In particular, it is possible to factor out the source contribution to the spectrum using secondary-to-primary ratios, such as deuteron-to-proton (d/p) and deuteron-to-helium-4 (d/$^4$He)~\cite{coste} to constrain the parameters of the galactic propagation model.

Isotope identification for singly charged particles has already been performed by magnetic spectrometers such as PAMELA~\cite{PAMELA_d}, IMAX~\cite{IMAX92}, and CAPRICE~\cite{CAPRICE98} for CR energies up to a few GeV/n. The Alpha Magnetic Spectrometer (AMS-02)~\cite{AMSPhysReport} will extend the energy range of isotopic composition measurement~\cite{Delgado:20230c} to $\sim$10 GeV/n. However, the analysis presents challenges due to the intrinsic characteristics of the measurement. Isotopes are separated through their mass by combining the rigidity ($R = p\,c/Z\,e$, momentum per unit charge) and the particle's velocity, as follows:
\begin{equation} \label{eq:mass}
m = \frac{R \, Z \, e}{\beta \, \gamma}\, ,
\end{equation}
where $Ze$ is the magnitude of the charge, $\beta = v/c$ is the velocity in speed of light units, and $\gamma$ is the Lorentz factor. The mass resolution can be derived from Equation~(\ref{eq:mass}):
\begin{equation}
\left(\frac{\Delta m}{m}\right)^2 = \left(\frac{\Delta R}{R}\right)^2 + \gamma^4 \left(\frac{\Delta \beta}{\beta}\right)^2\,.
\end{equation}

Due to the dependence on the fourth power of the Lorentz factor, the velocity resolution's contribution to the total mass resolution dominates for $\beta \rightarrow 1$, which happens for energies higher than a few GeV/n. Therefore, efficiently cleaning the initial sample from events whose RICH velocities have not been accurately reconstructed is essential to identify singly charged isotopes and extend the energy range of the measurement.

Artificial intelligence (AI) methods, particularly machine learning (ML), are widely used for particle identification in particle and astroparticle physics~\cite{miniboone, Graziani:2021vaiAMS, AMS:2014gdf, AMS:2014bun, AMS:2013fma, GRAZIANI20162351}. In particular, boosted decision trees (BDTs) have been employed to classify events with misreconstructed RICH velocities and reject the background for the identification of deuterons~\cite{Bueno:2023vrg}.
For this type of classification task, complex data sets containing many variables and parameters are analysed, and data are often noisy and contain correlated information, which is thus redundant. Hence, a crucial step in the application of ML methods to classification tasks is selecting the variables (or features) that constitute the input required for model training. ML-driven feature selection techniques are emerging as suitable tools to optimise the performance of ML algorithms for classification tasks in particle~\cite{DiLuca:2023byz} and astroparticle physics~\cite{Finke:2020gbv, Gavrikov:2021ktt, Luo:2020bbk}. Furthermore, feature selection has already been used in cosmic ray identification for ground-based experiments in~Herrera et al.~\cite{Herrera:2020twt} to rank the relevance of features involved in primary particle reconstruction from air shower simulations. The importance of feature selection lies in its ability to simplify the data analysis process. By identifying and selecting the most relevant features, it is possible to enhance the efficiency and accuracy of the classification algorithms, making the results more interpretable and robust. Moreover, feature selection aids in the avoidance of overfitting, a common pitfall in complex data sets. Focusing on essential features reduces the risk of models becoming overly tailored to the training data, thereby increasing their generalisation capabilities.
%As pointed out in~\cite{IceCube:2014slq}, this selection is a critical step in the process of background reduction analysis since it allows us to find the best set of features while reducing the computational power needed for the classification task.

%This study aims to analyse feature selection techniques applied to background reduction to identify singly-charged cosmic-ray isotopes with the AMS-02 detector.
{In this work, we study machine learning (ML) algorithms for feature selection in the context of singly charged cosmic ray isotope identification using the AMS-02 experiment and investigate whether these ML techniques could enhance the efficiency in reducing the RICH background compared to traditional physics-driven methods, such as~Bueno~et~al.~\cite{Bueno:2023vrg}.}
Five ML techniques are used to single out, from a sample of 130~features obtained by the detection and reconstruction of cosmic ray nuclei with the RICH detector of the AMS-02 experiment, the most promising features to identify the signal and reject the background. Furthermore, for comparison, the physics-driven approach proposed by~Bueno et al.~\cite{Bueno:2023vrg} is used to choose a set of features based on the knowledge of the detector and of the type of background to be reduced.

The paper is organised as follows: In Section~\ref{sec:RICH}, the AMS-02 detector is presented. The preparation of the data set and the algorithms used to perform feature selection are described in Section~\ref{sec:2}, along with the metrics used to evaluate the performance of the different methods. In Section~\ref{sec:3}, the results of the different models and their performance on the validation basis are followed by a discussion on the features selected by the best-performing model. We conclude in Section~\ref{sec:5}.

%=================================================================
%=================================================================
%=================================================================

\section{The AMS-02 RICH Detector}
\label{sec:RICH}

The Alpha Magnetic Spectrometer (AMS-02) is a cosmic ray detector operational aboard the International Space Station since May 2011~\cite{AMSPhysReport}. Its unique capabilities allow for the measurement of deuteron flux in previously unexplored energy ranges, extending it nowadays above a limit of 4 GeV/n.

AMS-02 consists of several subsystems as follows: A silicon tracker with nine layers, positioned from the top to the bottom of the detector, in conjunction with a permanent magnet of 0.15~T. This combination enables the measurement of the magnitude, the sign of the charge, and the rigidity of the particles; a transition radiation detector (TRD), designed to distinguish between leptons and hadrons; a time-of-flight (TOF) system, comprising two pairs of scintillators (upper TOF and lower TOF) located above and below the magnet, is responsible for measuring the velocity and charge $Z$ of the particles and serves as the primary trigger for the experiment; a ring-imaging Cherenkov detector (RICH), positioned below the lower TOF, is used to measure the particle velocity and charge $Z$; an anti-coincidence counter (ACC), which identifies and rejects particles with high-incidence angles; an electromagnetic calorimeter (ECAL), positioned  below the RICH, is responsible for measuring particle energy and enabling differentiation between leptons and hadrons.

\textls[-15]{The RICH detector~\cite{amsrich} plays a crucial role in measuring hydrogen isotope fluxes, allowing for their identification up to 10 GeV/n~\cite{efb}. However, it is essential to emphasise that this task presents significant challenges due to the intrinsic nature of the Cherenkov effect. Because the intensity of the emitted radiation is proportional to $Z^2$~\cite{jackson}, singly charged isotopes generate a faint signal compared to higher $Z$ particles, making the velocity reconstruction vulnerable to background disruptions, especially near the threshold of each~radiator.}

\begin{figure}[H]%[!h]
%    \centering
\includegraphics[width=0.5\linewidth,keepaspectratio]{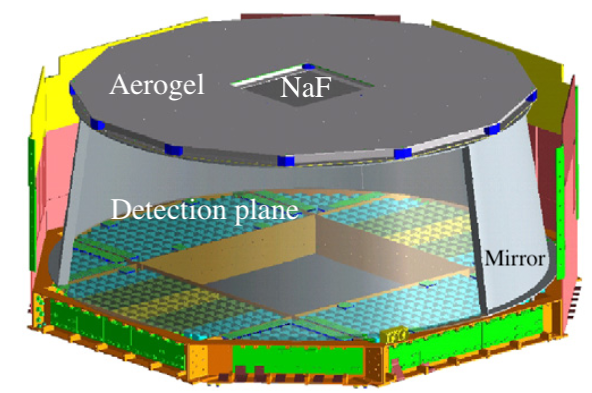}      %\includegraphics[width=0.6\linewidth,keepaspectratio]{rich.png}
\caption{Sketch of the ring-imaging Cherenkov detector of AMS-02. Adapted with permission from Ref.~\cite{AMSPhysReport}. Copyright 2011, AMS Collaboration.}
\label{fig:ams02rich_ch4}
\end{figure}

The RICH detector~\cite{Arruda:2007dhb, Aguilar-Benitez:2010xtr, Giovacchini:2020ilv}  (see Figure~\ref{fig:ams02rich_ch4}) features a truncated conical shape with a 60 cm top radius, a 67 cm bottom radius, and an expansion height of 47 cm. The detector comprises a radiator plane, an expansion volume, and a photo-detection plane. The double radiator plane includes a central radiator consisting of 16 tiles of sodium fluoride (NaF) measuring 8.5~×~8.5~×~0.5 cm$^3$, with a refraction index of 1.33. It is surrounded by 92 silica aerogel tiles measuring 11.5~×~11.5~×~2.5 cm$^3$ and having a refraction index of 1.05.

{The detection plane is equipped with an array of 680 photomultiplier tubes (PMTs) arranged in eight grids, four rectangular and four triangular. %The PMT model is the 4~×~4 multi-anode Hamamatsu R7600-00-M16, with a fast and uniform response, low operational voltage, and reduced sensitivity to external magnetic fields. Cherenkov photons are directed to the photocathode via an array of 10,880 solid pyramidal light guides made of Diakon LG-703, featuring different shapes to maximise photon collection efficiency. Consequently, a detection cell comprises a photomultiplier coupled to 16 light guide pipes bonded together, creating a continuous detection surface of $34 \times 34$ mm$^2$. The final spatial granularity of the detection plane is 8.5~×~8.5 mm$^2$, with 3 mm gaps between PMTs.
To minimise lateral losses of approximately $30$\% of the radiated Cherenkov photons, the expansion volume is surrounded by a highly reflective mirror that meets roughness specifications of better than 150~nm and exhibits a reflectivity exceeding $90$\% at $\lambda=420$ nm.
}
\vspace{-3pt}
  %MDPI: figures should be inserted after the it's first mention. Please check References should be numbered in order of appearance after modification.
%MDPI: Please ensure that permission has been obtained and there is no copyright issue. If copyright is needed, please provide a citation in the following format: "Reprinted/adapted with permission from Ref. [XX]. Copyright year, copyright owner's name". More details on "Copyright and Licensing" are available via the following link: https://www.mdpi.com/ethics#10.

%====================================================
%====================================================
%====================================================

\section{Methodology}
\label{sec:2}

\subsection{Database Description}
\label{sec:2.1}

%Six months of data collected by AMS-02 in space from December 9, 2015, to May 9, 2016, were used in this work.
{Six months of data collected by AMS-02 in space (from December 2015 to May 2016) were used in this work. While this particular choice is arbitrary, we believe it does not introduce any bias to our analysis, providing us with a statistically relevant sample. The performance of the AMS-02 detector has been extensively verified in the past years~\cite{rich_test} and no specific time-dependent effect on performance was reported.
%any contiguous six-month period could have been employed without compromising the validity of our analysis.
}

A first selection was applied to ensure the quality of the reconstructed tracker track and of the velocity measured by TOF for the used events. Singly charged isotopes were selected using TOF and silicon tracker charge measurements. The requirements on the tracker charge ($Z_{TRK}$) and upper and lower TOF charge ($Z_{UTOF}, Z_{LTOF}$) are $0.75 < Z_{TRK, UTOF}< 1.5$ and $0.75 < Z_{LTOF}< 1.3$, respectively. Then, the selected events were divided into two samples based on their reconstructed masses, following~Bueno et al.~\cite{Bueno:2023vrg}: events with a mass within 2$\sigma$ from the proton mass (0.75 GeV/c$^2<m<$ 1.25 GeV/c$^2$) or a mass above $4\sigma$ from the triton mass ($m>$ 4 GeV/c$^2$) are considered to be signal-like or background-like, respectively. This preparation is necessary to have a labelled set of events to be used for the training of the classification method.
%This choice is based on the fact that the background for the identification of deuterons, composed by particles whose velocity has been misreconstructed, predominantly comprises events in the high mass tail of the particles' mass distribution. Conversely, the chosen mass range for the signal-like sample corresponds to the proton peak of the mass distribution and it is characterized by particles consistent with signal events, in other words events whose velocity has been correctly reconstructed.

As was pointed out in~Bueno et al~\cite{Bueno:2023vrg}, the residual background for the identification of cosmic ray deuterons consists mainly of events whose RICH velocity is poorly reconstructed due to noise disrupting the already weak signal produced by the $Z=1$ particles. In particular, particles produced from the interactions of the incoming cosmic rays with the AMS-02 detector can induce spurious hits that are not related to the Cherenkov emission of cosmic rays. These ring-uncorrelated hits consist of additional photons generated as the particles produced by the interactions in the detector cross the aerogel radiator or while the same particles cross the PMT plane. These spurious events induce additional photon hits and consequently affect the reconstruction of the Cherenkov ring. Furthermore, interactions between particles happening in the region between the lower tracker and the RICH can modify the direction of the incoming cosmic ray, thus introducing a slight bias in the number of detected photon hits in the Cherenkov ring with respect to the reconstructed tracker track. These events whose velocities are poorly reconstructed constitute the background of this analysis and they are the ones to be rejected to improve mass resolution and correctly identify singly charged isotopes. They are mainly located in the high mass tail of the particles' mass distribution, as outlined in~Bueno et al.~\cite{Bueno:2023vrg}; for this reason, the background-like sample comprises events falling within that specific mass distribution region. Conversely, the mass range selected for the signal-like sample corresponds to the proton peak of the mass distribution and it is characterised by events whose velocity has been correctly reconstructed.

\begin{figure}[H]%[ht]
\includegraphics[width=13.7 cm]{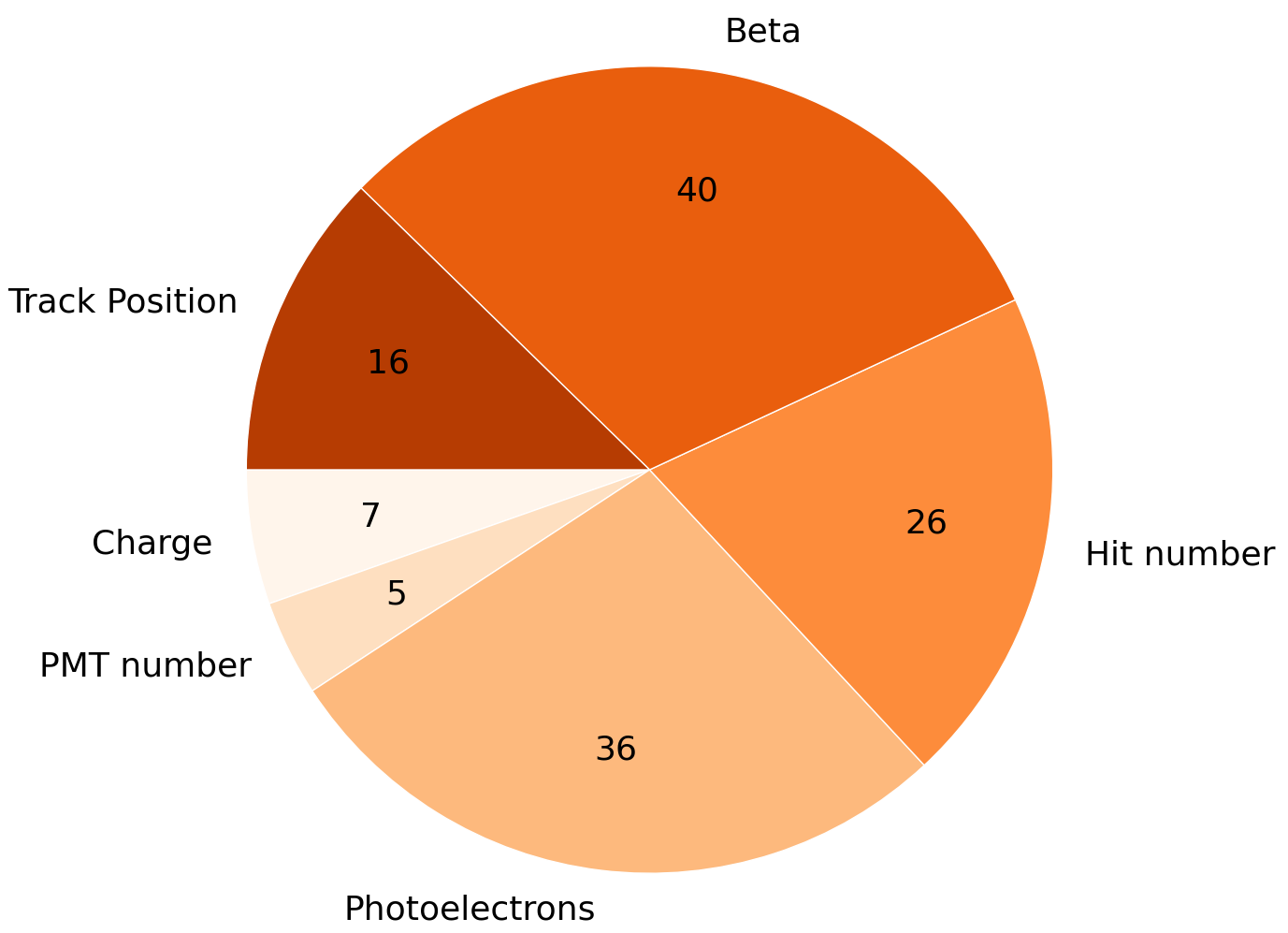}
\centering
\caption{Pie chart showing the breakdown of the 130 features into six classes. The value in each slice represents the number of features in the corresponding class.} %Attention AE: 'Position' in the figure would be better as 'position' (lower case p) for consistency with rest of figure.

\label{piechart_complete}
\end{figure}

A set of 130 features was used for the analysis, including quantities directly measured by the RICH detector (e.g., the number of crossed PMTs and the number of hits), the reconstructed quantities (e.g., charge and velocity), and the expected values of the measured and reconstructed features computed using the reconstruction and input parameters of the event itself (e.g., expected number of photoelectrons associated with the Cherenkov ring). As shown in Figure~\ref{piechart_complete}, the used features can be divided into six classes: Charge, Track position, PMT number, Beta, Hit number, and Photoelectrons.

The class ``Charge'' contains seven features related to the magnitude of the reconstructed charge: these include two different estimates of the charge based on different Cherenkov ring reconstruction methods~\cite{barao_rich,Delgado_rich}, the expected charge resolution and its mean square error, and the value for the Kolmogorov probability associated to the hypothesis of a uniform distribution of signal hits along the particle path~\cite{1971smep.book.....E}.
The class ``Track Position'' contains 16 features related to the extrapolated tracker track of the particle inside the RICH, such as the coordinates of the impact point on the radiator plane and the angles that the reconstructed track forms with it, the distance of the impact point from the border of the radiator tile, and the information on the radiator tile crossed by the extrapolated track.
The class ``Beta'' contains 40 features connected to the reconstructed velocity of the particle: it includes, for example, the particles' velocity, reconstructed independently by two methods~\cite{barao_rich,Delgado_rich}, the values of the velocity resolution, and the mean square error. %, and the average velocity and resolution for the first 10 hit clusters ordered by probability.
The class ``Hit number'' delivers information about the number of photons detected in the photodetection plane by means of the number of photoelectron hits registered by the PMTs. Among these 26 features are included the total number of detected hits, and the number of hits inside and outside the Cherenkov ring. %and the number of hits compatible with reflected photons or compatible with a particle with a beta equal to one.
The class ``Photoelectrons'' includes 36 features incorporating different information about the measured number of photoelectrons (p.e.) deriving from the detection of the Cherenkov photons in the PMTs of the detection plane. For every event, the number of p.e. in the PMT with the highest number of p.e., the number of p.e. collected in and out of the ring, and the number of p.e. expected for a singly charged nucleus or an electron with the reconstruction and input parameters of the current event are available.
Finally the class ``PMT number'' is related to the number of PMTs crossed in each event, and includes five features related to the measured and expected number of PMTs inside and outside of the Cherenkov ring. {All the variables used in the analysis are listed and briefly described in Table~\ref{tab:features_list}.}

\vspace{-4pt}

\begin{figure}[H]%[]
\begin{adjustwidth}{-\extralength}{0cm}
\centering
\includegraphics[width=15.5cm]{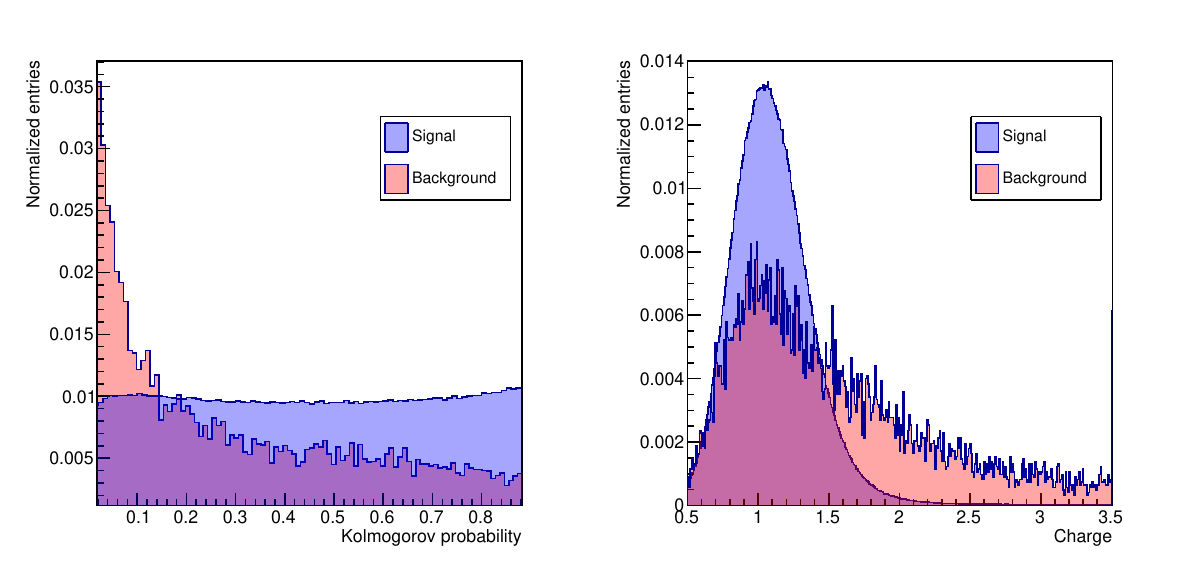}
\end{adjustwidth}
\caption{The distributions %MDPI: We moved this figure after where it is first mentioned in the text. Please confirm.
of reconstructed charge (\textbf{left}) and Kolmogorov probability (\textbf{right}) for the aerogel radiator for the signal-like (in blue) and background-like samples (in red). The purple regions represent the overlaps between these two samples.}
\label{feat_ex}
\end{figure}

To illustrate the different behaviour that signal- and background-like events display, Figure~\ref{feat_ex} shows the distributions of the Kolmogorov probability and the magnitude of the charge, denoted by $Z$, for a signal-like (in blue) and a background-like (in red) sample of events. Both features show different distributions for the two samples; hence, they discriminate well between background-like and signal-like events. This conclusion derives from the physical phenomena that these features trace.
The left plot shows the distribution of the Kolmogorov probability~\cite{1971smep.book.....E} that is obtained by performing for each event a Kolmogorov test on the azimuth distribution of emitted photons along the particle path,  expected to be uniform for well-reconstructed events with a ring-like shape (i.e., signal-like events), and to be non-uniform for background events, whose rings include noisy hits.
The Kolmogorov test was used to compare the expected cumulative distribution for the azimuth angle with the measured one. When the discrepancy between these two distributions is maximal the Kolmogorov probability scores very low values, justifying the behaviour of the background-like sample in the left plot in Figure~\ref{feat_ex}.  On the other hand, the Kolmogorov probability is calculated such that it is almost uniformly distributed between $0$ and $1$ for ring-shaped events, as shown by the signal-like sample in the same plot.
The distribution of the reconstructed charge is shown in the right panel of Figure~\ref{feat_ex}. In the RICH detector, the square of the charge $Z$
%MDPI: Please confirm that all variables are in uniform format. (regular/italic/bold).
is proportional to the number of photoelectrons associated with the Cherenkov ring (i.e., $Z^2 \propto N_{p.e.}$), with a scaling factor accounting for the ring acceptance and velocity dependence. We expect signal-like events to have a symmetric charge distribution peaked at $Z=1$, as in Figure~\ref{feat_ex}. Conversely, the background-like sample mostly contains events whose rings have spurious hits, leading to the characteristic high charge tail in the red distribution.
%In the RICH the charge $Z$ is given by $Z = \sqrt(N_{p.e.}/N_{exp})$, where $N_{p.e.}$ is the number of photoelectrons associated with the ring and $N_{exp}$ is the number of photoelectrons expected for a $Z=1$ particle with the same RICH velocity and tracker track of the current event. Since a signal-like event is required to have a value of $N_{p.e.}$ very similar to $N_{exp}$, we expect signal-like events to have a symmetric charge distribution peaked at $Z=1$, as in Figure~\ref{feat_ex}. Conversely, the background-like sample mostly contains events whose Cherenkov rings are disrupted by noise and hence badly reconstructed, leading to the characteristic high charge tail in the red distribution.

\vspace{-12pt}

%====================================================
%====================================================
%====================================================

\subsection{Feature Selection Techniques}
\label{sec:2.2}

Feature selection techniques play a crucial role in the data analysis and the modelling of ML algorithms. The importance of these techniques resides in the fact that not all attributes contribute equally to constructing an accurate and efficient model~\cite{bommert2020benchmark}. By selecting the relevant characteristics of the data set, it is possible to improve the generalisation capacity of the model, reducing the chance of overfitting~\cite{bommert2020benchmark, dvornik2020selecting, han2021hybrid}. Furthermore, other benefits obtained when applying feature selection are the reduction in noise and interference and the acceleration of the training time and computational efficiency, as it reduces processing resources~\cite{jia2022feature, qian2023survey}. Therefore, feature selection techniques are essential in optimising and improving ML models, allowing better interpretability, efficiency, and accuracy of data~analysis.

The most straightforward approach {in selecting the relevant features} is to test every possible subset of features, finding the one that minimises the error rate. However, this is an exhaustive and computationally intractable search for real data sets. In this sense, the choice of the evaluation metric strongly influences the feature selection technique, and it is these evaluation metrics that distinguish between the three main categories of feature selection techniques
~\cite{chen2020ensemble, effrosynidis2021evaluation}, namely:

\begin{itemize}

\item Filter: Use a proxy measure instead of an error rate to score a subset of features;

\item Wrapper: Use a predictive model to score subsets of features. Each new subset is used to train a model, which is tested on a validation set;

\item Embedded: Is a comprehensive group of techniques that perform feature selection as part of the model-building process.

\end{itemize}

Thus, in order to achieve the project's objective of verifying the most relevant features in the database, we selected a technique for each category of feature selection techniques, namely: SelectKBest (filter), Random Forest - RF (wrapper), and linear regression (embedded). In addition to the mentioned techniques, Pearson's correlation is applied as a feature selection technique. We briefly describe the techniques used:

\begin{itemize}

\item Kbest: is an approach that selects the \textit{k} best attributes based on a statistical measure, such as the analysis of variance (ANOVA) used in this study. By defining a value for \textit{k}, it is possible to choose the \textit{k} most significant features, which have a more relevant impact on the model's prediction~\cite{liu2020feature, alves2023automated};

\item Random Forest (RF): is a technique that can be applied not only for building classification and regression models but also as a practical feature selection tool~\cite{genuer2010variable}. Random Forest performs several independent decision trees, each using different subsets of attributes and random samples from the data set. During this process, the algorithm calculates the importance of each attribute based on its contribution to the overall accuracy of the predictions~\cite{effrosynidis2021evaluation};

\item Linear Regression: the approach considers the coefficients of each attribute as a measure of the individual contribution in predicting the dependent variable~\cite{polat2020detecting}. The magnitude and sign of the coefficient indicate the impact on the target variable. Thus, features with higher and statistically significant coefficients are considered more important and can be selected as part of the feature selection process~\cite{effrosynidis2021evaluation, polat2020detecting};

\item Pearson's Correlation (CORR): the approach measures the strength and direction of the linear relationship between two continuous variables~\cite{seeram2019overview}. When calculating the Pearson correlation between each attribute and the target features, it is possible to obtain a value ranging from $-$1 to 1. A value close to $-$1 indicates a strong negative correlation. On the other hand, a value close to 1 indicates a strong positive correlation. A value close to 0 indicates a weak or no correlation. Based on the correlation values, it is possible to select attributes with a more significant correlation with the target variable~\cite{bommert2020benchmark, liu2020daily}.

\end{itemize}

Furthermore, for comparison purposes, the methodology proposed by~Bueno et al.~\cite{Bueno:2023vrg} is considered to select the last set of features. This method involves the detailed study of the events whose mass is incorrectly identified due to the interactions occurring within the AMS-02 detector and aims to identify the sources of interactions relevant to the RICH reconstruction background and mitigate this background efficiently. In this work, the features are chosen following a physics-driven approach based on the knowledge of the RICH detection mechanism and velocity reconstruction method and they can be combined to form more effective features. A multivariate estimator is subsequently used to complete the classification task.

%In particle identification, Boosted Decision Trees (BDTs) are well-established ML techniques that are widely used in particle and astroparticle physics~\cite{bdt, Coadou2022}. In particular, they have been successfully employed in the frame of the AMS-02 Collaboration especially to identify cosmic-ray electrons and positrons while rejecting protons~\cite{aguilar2013first, aguilar2014precision, ams2014high, graziani2016electron}, based on the analysis of the three-dimensional reconstruction of the shower shape generated by the Electromagnetic Calorimeter (ECAL). In addition, BDTs have proven to be particularly suitable for particle identification with the RICH detector, due to their ability to incorporate multiple features associated with Cherenkov ring reconstruction~\cite{Bueno:2023vrg}, thus enabling the distinction of events with different degrees of accuracy in velocity reconstruction.

\subsection{Performance Evaluation Metrics}
\label{sec:2.3}

Accurately evaluating the performance of the algorithms is essential to guide the choice of the best classification model~\cite{hossin2015review}. Metrics such as accuracy, precision, F1-score, and recall play a central role in this evaluation, allowing the effectiveness and usefulness of the algorithms to be measured~\cite{yacouby2020probabilistic}. In the following, we will explore the meaning and calculation of these metrics, highlighting their relevance and providing a general description to ensure reliable and accurate AI systems:

\begin{itemize}

\item Accuracy: this metric provides a general measure of the model's ability to correctly predict classes. It is helpful in scenarios where all classes have similar importance. It considers true positives (correctly classified cases) and true negatives (correctly classified negative cases) concerning the total number of examples. Accuracy is defined as follows:
\begin{equation}
\text{Accuracy} = \frac{\text{True Positives} + \text{True Negatives}}{\text{Total Examples}};
\end{equation}

\item Precision: this metric focuses on the quality of the model's positive predictions. It is particularly relevant when false positives have a substantially more significant impact than false negatives. It provides deeper insight into the model's ability to avoid the erroneous classification of negative examples as positive. Precision is defined as~follows:
\begin{equation}
\text{Precision} = \frac{\text{True Positives}}{\text{True Positives} + \text{False Positives}};
\end{equation}

\item Recall: this metric, also known as sensitivity, focuses on the model's ability to identify positive cases while effectively minimising false negatives. The recall metric is defined as follows:
\begin{equation}
\text{Recall} = \frac{\text{True Positives}}{\text{True Positives} + \text{False Negatives}};
\end{equation}

\item F1-score: this metric combines the precision and recall metrics to provide a balanced measure of model performance. It is particularly relevant when the balance between accurately identifying positive cases and minimising false positives and false negatives is essential. The F1-score is defined as follows:
\begin{equation}
\text{F1-score} = 2 \times \left( \frac{\text{Precision} \times \text{Recall}}{\text{Precision} + \text{Recall}} \right);
\end{equation}

\end{itemize}

In summary, performance evaluation metrics, including accuracy, precision, recall, and F1-score, play an essential role in evaluating classification algorithms. They provide valuable insights into the quality and effectiveness of forecasts, adapting to different needs and contexts.

%====================================================
%====================================================
%====================================================

\section{Experiments, Results, and Discussion}
\label{sec:3}

This section describes the experiments conducted in this study and provides a critical evaluation of the results obtained. Section~\ref{sec:3.1} describes the steps involved in processing and balancing data to create an equitable and reliable data set. Next, Section~\ref{sec:3.2} examines the strategies employed to identify the most informative features of data sets. Subsequently, Section~\ref{sec:3.3} presents the results achieved through the developed models, highlighting their performance metrics and predictive capacity. Finally, Section~\ref{sec:3.4} explores the role and contribution of the Random Forest algorithm in achieving the study objectives. %The focus of this section is to provide a detailed and reasoned view of the procedures and results, to elucidate the study's conclusions, and to contribute to the advancement of knowledge in the study context.

\subsection{Preparation of the Database for the Experiments}
\label{sec:3.1}

%When conducting feature selection experiments in analyzing unbalanced data sets, we used the technique called ``RandomUnderSampler''~\footnote{RandomUnderSampler performs a random and strategic selection of samples from the majority class, reducing their representativeness and, thus, levelling the database for subsequent analyses.}. This methodology balances the original database, reducing the disparity between the interest categories (signal and background). Such an imbalance can introduce significant biases in statistical analysis and modelling, undermining the effectiveness and reliability of machine learning algorithms. This methodology allows feature selection algorithms to operate in a more balanced context, promoting the fair evaluation of features regardless of the class to which they are associated. As a result of this procedure, a new data set was obtained consisting of 33.234 samples, 16.617 of which were signal and 16.617 were background.

The data sample used in this work is made of almost 3 million well-reconstructed singly charged events crossing the RICH detector of AMS-02. This sample primarily comprises events whose mass is well reconstructed (signal), with a tiny fraction of about one percent of events whose mass is misreconstructed (background). The disproportion between the two samples can introduce significant biases in statistical analysis and modelling, undermining the effectiveness and reliability of ML algorithms. To solve this issue, we use the \texttt{RandomUnderSampler} technique available on the \texttt{Imbalanced-learn} Python toolbox~\cite{JMLR:v18:16-365}, which performs a random and strategic selection of samples from the majority class, reducing their representativeness and, thus, levelling the data set for subsequent analyses, lowering the disparity between the interest categories (signal and background). As a result of this balancing procedure, a final data set consists of 33.234 events, 16.617 of which are signal and 16.617 background.

%In the subsequent stage, after balancing, the data set was divided into two sets to perform the classification task: training data and validation data, using the method called "train\_test\_split", enabling the development and optimisation of learning models machine in a controlled environment. The training data set (70\% of the base) is intended for selecting the most relevant characteristics using feature selection techniques. In comparison, the validation data set (30\% of the base) is reserved for the evaluation and validation of the results obtained.

In the subsequent stage, after balancing, the data set was divided into two sets to perform the classification task: training data and validation data. The training data set (70\% of the events) is intended for selecting the most relevant characteristics using feature selection techniques. In comparison, the validation data set (30\% of the events) is reserved for the evaluation and validation of the results obtained.

%In this study the Boosted Decision Tree algorithm (AdaBoostClassifier) was used, with the parameter number of estimators ($n\_estimators$) set at 100.

The described approach is followed for obtaining solid and reliable models capable of dealing with unbalanced data, selecting the most informative features, and validating their effectiveness in independent test environments, thus contributing to more accurate analyses and informed decisions.

{Table~\ref{tab:parameters} details the parameters used in the employed feature selection techniques, which include Kbest, Random Forest, linear regression, and correlation. We chose to use the default values of the respective software libraries to ensure reproducible results.}

\begin{figure}[H]%[h!]
\begin{adjustwidth}{-\extralength}{0cm}
\centering
\includegraphics[width=18cm]{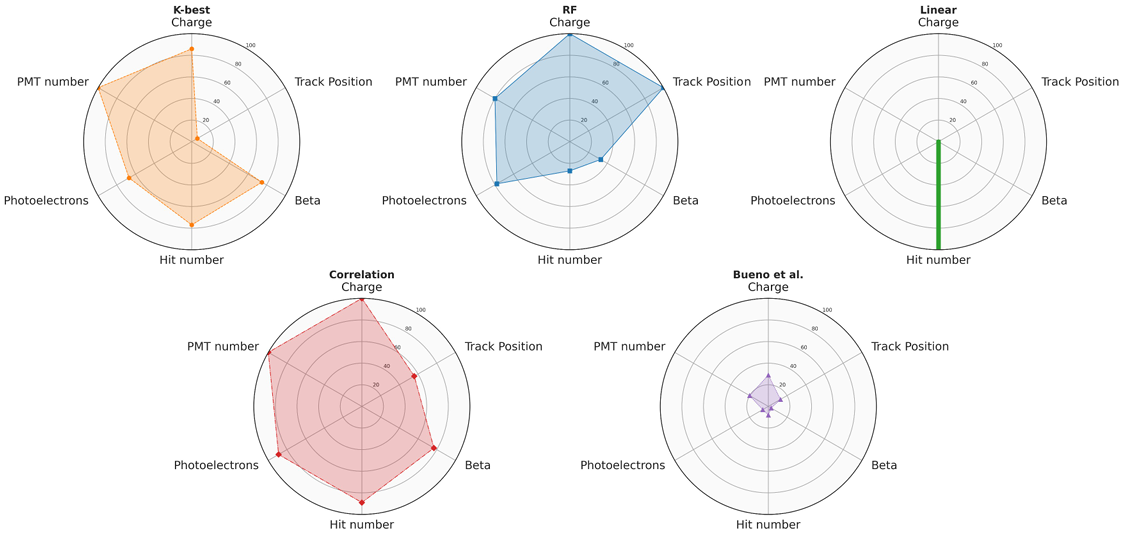}
\end{adjustwidth}
\caption{Fraction of features selected by each method out of the six classes discussed in this work. See text for discussion.}
\label{piechart:Radar_01}
\end{figure}

{Figure~\ref{piechart:Radar_01} presents a visual comparison between the selected methods for the identified classes (Charge, Track Position, Beta, Hit number, Photoelectrons, and PMT number). Each method is represented in an individual graph:
%visually displaying the feature classes selected by each technique.
the values on the radial axes represent the percentage of selected features for each class in relation to the number of features originally contained in the same class, allowing a quick and comprehensive comparative analysis of the selected features of each method in the different classes. Furthermore, Table~\ref{tab:percentagesone} complements this visual representation by listing the number (and percentages) of the features selected for each class by each technique.}

\vspace{-12pt}

\begin{table}[H]%[h!]
\caption{Number (and percentages) of the feature selected for each class by the different methods}
\begin{adjustwidth}{-\extralength}{0cm}
\newcolumntype{C}{>{\centering\arraybackslash}X}
\begin{tabularx}{\fulllength}{CCCCCCCC}
\toprule
& \textbf{Charge}    & \textbf{Track Position}   & \textbf{Beta}        & \textbf{Hit Number}  & \textbf{Photoelectrons} & \textbf{PMT Number} & \textbf{Total} \\ \midrule
Kbest       & {6 (86\%)}  & {1 (6\%)}  & {30 (75\%)}   & {20 (77\%)} & {24 (67\%)}    & {5 (100\%)} & 86  \\
RF          & {7 (100\%)} & {16 (100\%)} & {13 (33\%)} & {7 (27\%)}  & {28 (78\%)}    & {4 (80\%)} & 75  \\
Linear      & {0 (0\%)}   & {0 (0\%)}    & {0 (0\%)}     & {1 (100\%)}   & {0 (0\%)}        & {0 (0\%)} & 1   \\
Correlation & {7 (100\%)} & {9 (56\%)} & {31 (77\%)} & {23 (89\%)} & {32 (89\%)}   & {5 (100\%)} & 107  \\
Bueno et al. %MDPI: Please confirm if it is reference citation and check it in the full text.
& {2 (29\%)} & {2 (13\%)} & {1 (3\%)} & {2 (8\%)} & {2 (6\%)} & {1 (20\%)} & 9 \\
\bottomrule
\end{tabularx}
\end{adjustwidth}
\noindent
\label{tab:percentagesone}
\end{table}

%The physics-driven method selects the same percentage of features in each class, while
The percentage of variables selected by the ML methods for each class strongly depends on the algorithm itself. Nonetheless, there are some classes with a high percentage of features selected for all the methods used: this hints towards a connection between the discrimination power of the features and the physics underlying the detection mechanism, as will be discussed more in detail in Section~\ref{sec:3.4}.
%For comparing the selected feature selection techniques, all the variables in the database are used; the name used for this sample is ``All''.

%====================================================
%====================================================
%====================================================

\subsection{Analysis of Feature Selection Techniques}
\label{sec:3.2}

\begin{figure}[H]%[h!]
\begin{adjustwidth}{-\extralength}{0cm}
\centering
\includegraphics[width=17cm]{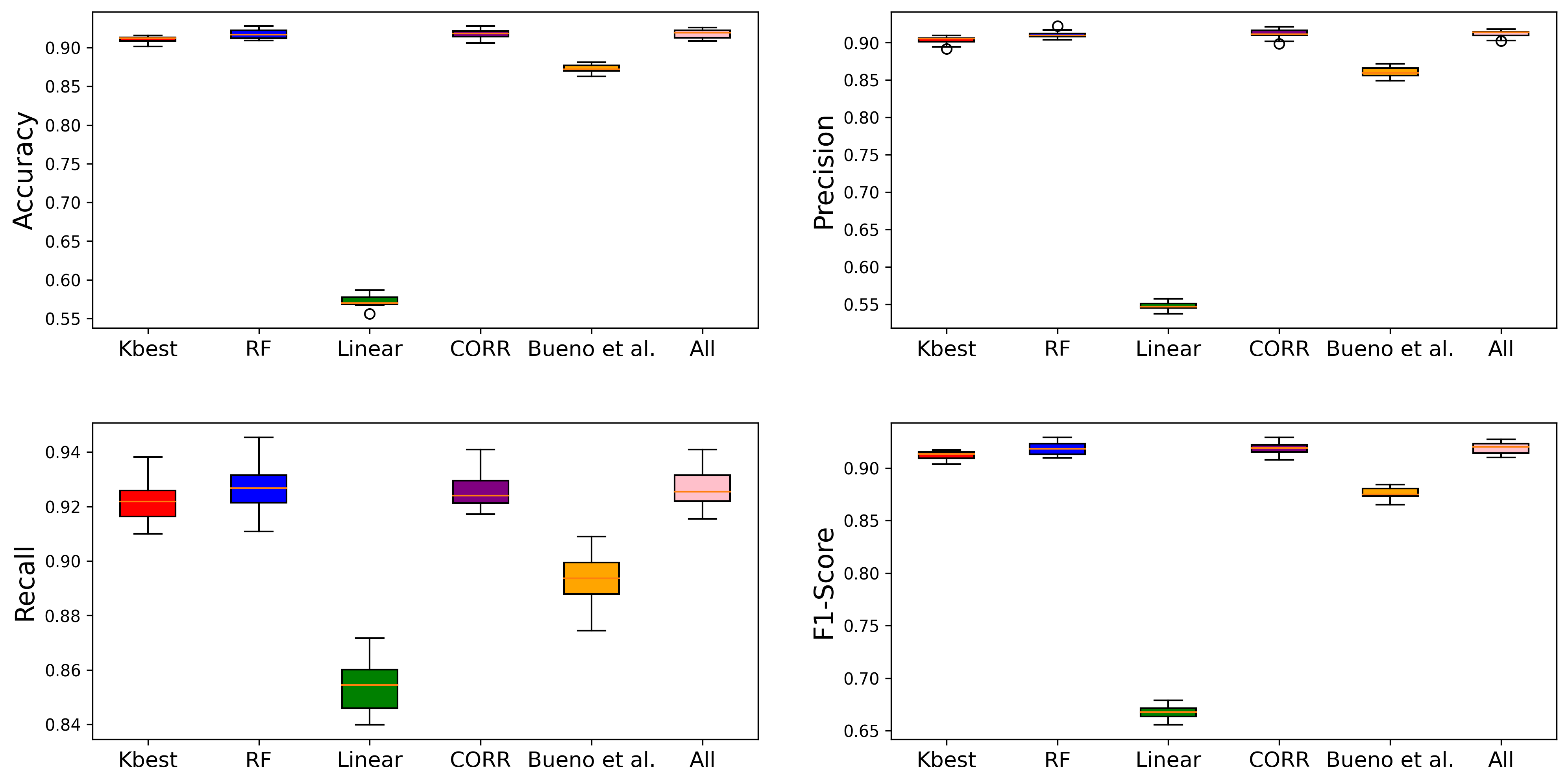}
\end{adjustwidth}
\caption{Boxplots showing the accuracy, precision, F1-score, and recall for the different selection algorithms used.}
\label{fig:boxplots} %  %MDPI: Please confirm if `Bueno et al.' in the images are reference citations.  %MDPI: Please ensure that permission has been obtained and there is no copyright issue. If copyright is needed, please provide a citation in the following format: "Reprinted/adapted with permission from Ref. [XX]. Copyright year, copyright owner's name". More details on "Copyright and Licensing" are available via the following link: https://www.mdpi.com/ethics#10.

\end{figure}

In this study, a boosted decision tree algorithm (AdaBoostClassifier) was used, with the parameter number of estimators ($n\_estimators$) set at 100.
%Initially, for training and evaluating the models, the performance of the feature selection techniques obtained through the Experimental Planning and Evaluation technique was evaluated;
In addition, \textit{k}-fold cross-validation was used, for training and evaluating the models,
with \textit{k} = 10, with \mbox{\textit{k} $-$ 1} for training and the rest for testing~\cite{mano2018emotional, mano2020intelligent}, %Thus, it is possible to measure the error estimate more accurately, as the average value estimate tends towards a valid zero error rate as \textit{n} increases
thus obtaining the average accuracy rate for each algorithm. The four panels of Figure~\ref{fig:boxplots} show the performance of the classification algorithm trained with the five sets of features obtained with the methods discussed in Section~\ref{sec:2.2}. For comparison, the performance of the classifier trained with all the variables in the database is also computed and labelled as ``All''.
%the Boxplots referring to the classifications of the feature selection techniques performed.
Table~\ref{tab:metrics_geral} shows the evaluation metrics (accuracy, precision, F-1 score, and recall) together with the \textit{p}-value (Shapiro--Wilks test) for the feature selection techniques used in this work. {The number of significant digits has been chosen based on the uncertainties reported in Table~\ref{tab:uncertainties}. %P-values are reported with the same number of significant figures for consistency.
}
%The performance of the algorithm is also evaluated when using all the variables in the database. This configuration is dubbed as ``All''.

\vspace{-12pt}

\newcommand{\mc}[3]{\multicolumn{#1}{#2}{#3}}
\begin{table}[H]%[!h]
\caption{Mean value of assessment metrics and \emph{p}-values for the sets of selected features.}
%\begin{center}
\label{tab:metrics_geral}
%\begin{tabular}{ccccccc}\hline
\begin{adjustwidth}{-\extralength}{0cm}
\newcolumntype{C}{>{\centering\arraybackslash}X}
\begin{tabularx}{\fulllength}{CCCCCC}
\toprule
\textbf{} & \textbf{Accuracy} & \textbf{Precision} & \textbf{F-1 Score} & \textbf{Recall} & \textbf{\textit{p}-Values (\textit{Shapiro--Wilk})}\\\midrule
Kbest & {0.911} & {0.903} & {0.912} & {0.923} & {0.860}\\
RF & {0.918} & {0.911} & {0.912} & {0.927} & {0.871}\\
Linear & {0.572} & {0.548} & {0.668} & {0.854} & {0.792}\\
Correlation & {0.918} & {0.911} & {0.919} & {0.927} & {0.063}\\
Bueno et al. & {0.873} & {0.860} & {0.876} & {0.893} & {0.977}\\
All & {0.918} & {0.911} & {0.920} & {0.927} & {0.395}\\
\bottomrule
%\end{tabular}
%\end{center}
\end{tabularx}
\end{adjustwidth}
\end{table}

\textls[-15]{{The Shapiro--Wilk test was applied to our data set to verify the normality hypothesis and, therefore, determine suitability for parametric or non-parametric tests.The \mbox{Shapiro--Wilk} test is a general test designed to detect all deviations from normality. The test rejects the hypothesis of normality when the \textit{p}-value is less than or equal to 0.05. Failing the normality test allows to assert with 95\% confidence that the data do not fit the normal distribution. %Passing the normality test will only enable to declare that no significant deviations from normality were found.)
%MDPI: Footnote is not permitted in our journal. Please include this paragraph to the maintext.
All techniques yield \emph{p}-values above the 0.05 threshold (see Table~\ref{tab:metrics_geral}),  corroborating the hypothesis of normal distribution of our data. However, a closer inspection reveals differences between their \emph{p}-values, highlighting distinct distribution profiles for each feature selection technique, and underscoring the importance of considering both statistical normality and performance metrics in the context of the AMS-02 experiment.}
%Notably, the Correlation method's p-value of 0.063, marginally exceeds the threshold, %suggests a potential deviation from normality not as pronounced in t
%while combining all the features (All technique), shows a p-value of 0.395. %This contrast may have implications for the robustness of feature selection under diverse conditions.
%Conversely, the Linear technique exhibits a high p-value of 0.792, indicating strong adherence to normality. Yet, its lower performance metrics call into question its performance in isotope identification with the AMS-02 experiment. The technique by~\hl{Bueno et al.}~\cite{Bueno:2023vrg} stands out with a p-value of 0.977, indicating high consistency in its results, although it does not show the best performance. %RF and Kbest, while similar in performance, show distinct p-values of 0.871 and 0.86, respectively. %Such differences, although minor, could become more pronounced under varied experimental conditions or with larger datasets.
%In summary, the Shapiro-Wilk test results corroborate the normal distribution of our data, yet the relative p-value differences highlight distinct distribution profiles for each feature selection technique, underscoring the importance of considering both statistical normality and performance metrics in the context of the AMS-02 experiment.
%As all \textit{p}-values obtained are more significant than 0.05 (see Table~\ref{tab:metrics_geral}), the normality hypothesis was accepted with 95\% confidence. Therefore, the parametric test is the most suitable for further analysis.
}

For the parametric test, the \emph{t}-test was used. The \emph{t}-test is a hypothesis test that uses statistical concepts to reject or not reject a null hypothesis. This assumption is usually accepted when the test statistic follows a normal distribution.  %MDPI: Footnote is not permitted in our journal. Please include this paragraph to the maintext.
This parametric method allows excellent reliability in data analysis with conformity to the normal distribution. Paired comparisons with the \emph{t}-test are shown in Table~\ref{tab:t-test}, and values less than 0.05 indicate a statistically significant difference between group results.

{
\begin{table}[H]%[!h]
\caption{\textit{p}-values of the pairwise comparison performed with the \emph{t}-test for the feature selection techniques.}
%\begin{center}
\label{tab:t-test}
\begin{adjustwidth}{-\extralength}{0cm}
\newcolumntype{C}{>{\centering\arraybackslash}X}
\begin{tabularx}{\fulllength}{CCCCCC}
\toprule
\mc{1}{c}{} & \textbf{Kbest} & \textbf{RF} & \textbf{Linear} & \textbf{Correlation} &\textbf{ Bueno et al.}\\
\midrule
\mc{1}{c}{RF} & {0.001} & - & - & - & - \\
\mc{1}{c}{Linear} & {0.000} & {0.000} & - & - & - \\
\mc{1}{c}{Correlation} & {0.027} & \textbf{0.704} & {0.000}  & - & - \\
\mc{1}{c}{Bueno et al.} & {0.000} & {0.000} & {0.000} & {0.000}  & - \\
\mc{1}{c}{All} & {0.002} & \textbf{1.000} & {0.000} & \textbf{0.658} & {0.000} \\
\bottomrule
\end{tabularx}
\end{adjustwidth}
%\end{center}
\end{table}
}

The results demonstrate that the features selected by the RF (75 features) and correlation (107 features) techniques do not present statistical differences if all 130 features are used since comparisons between pairs of results are more significant than 0.05. Therefore, only the RF and correlation approaches were considered for the following tests and compared to the approach including all features.

%====================================================
%====================================================
%====================================================

\subsection{Quantitative Analysis of Computational Complexity}
\label{sec:CPU_time}

{As shown in the previous section, the RF and correlation techniques do not show statistically significant differences compared to using the set including all variables. Hence, it is crucial to understand how these methods differ in processing time, which can be a limiting factor in practical applications. In this section, we present a quantitative analysis of computational complexity, specifically focused on the CPU time required to train a BDT. We compare the performance of the RF, of the correlation algorithm, and of the approach that employs all 130 features (All). This analysis is carried out on a computer equipped with an 11th Gen Intel(R) Core(TM) i9-11900KF @ 3.50GHz, 3.40 GHz processor to provide a solid basis for evaluation.}

{Table~\ref{tab:CPU_time} %MDPI: should it be `Table 4''?
presents the average CPU times used for each technique and their standard deviation, based on 100 runs, offering a detailed perspective on the variability and consistency between runs. We observe that the time required to train the BDT using the RF, correlation, and all techniques are, respectively, 659, 644, and 718 s.}

{
\begin{table}[H]%[!h]
\caption{{Average processing time for the RF, and correlation techniques, compared with the performance for all features.}}
%\begin{center}
\label{tab:CPU_time}
\newcolumntype{C}{>{\centering\arraybackslash}X}
\begin{tabularx}{\textwidth}{CCC}
\toprule
\textbf{} & \textbf{Average Time [s]} & \textbf{Standard Deviation [s]} \\ \midrule
RF &  659 & 8.3 \\
Correlation & 644 & 12.6 \\
All & 718 & 9.4 \\
\bottomrule
\end{tabularx}
%\end{center}
\end{table}
}

%These results are notable as they do not reflect significant statistical differences in terms of performance metrics, as seen in the previous section.
%Even if the three methods considered do not show significant statistical differences in terms of performance metrics, they are characterised by slightly different CPU times.
{The similarity in processing times among the three techniques indicates that, despite minor differences, these variations do not result in a notable advantage or disadvantage in terms of time performance. In essence, selecting a technique based solely on processing time may not be a decisive factor, as all methods are relatively efficient. Therefore, the choice between the RF, correlation, and all techniques should be based on a comprehensive assessment that considers performance both in terms of evaluation metrics and computational efficiency.}

%\textcolor{blue}{Therefore, the choice between the RF, Correlation, and All techniques should be based on a holistic assessment that considers both performance in terms of evaluation metrics and computational efficiency. Given that the differences in CPU time are minimal and there are no significant disparities in the evaluation metrics, the decision should be guided by additional criteria relevant to the specific application context.}

%====================================================
%====================================================
%====================================================

\subsection{Assessment of Predictive Models}
\label{sec:3.3}

\begin{figure}[H]%[!h]
%    \centering
\includegraphics[width=0.7\linewidth,keepaspectratio]{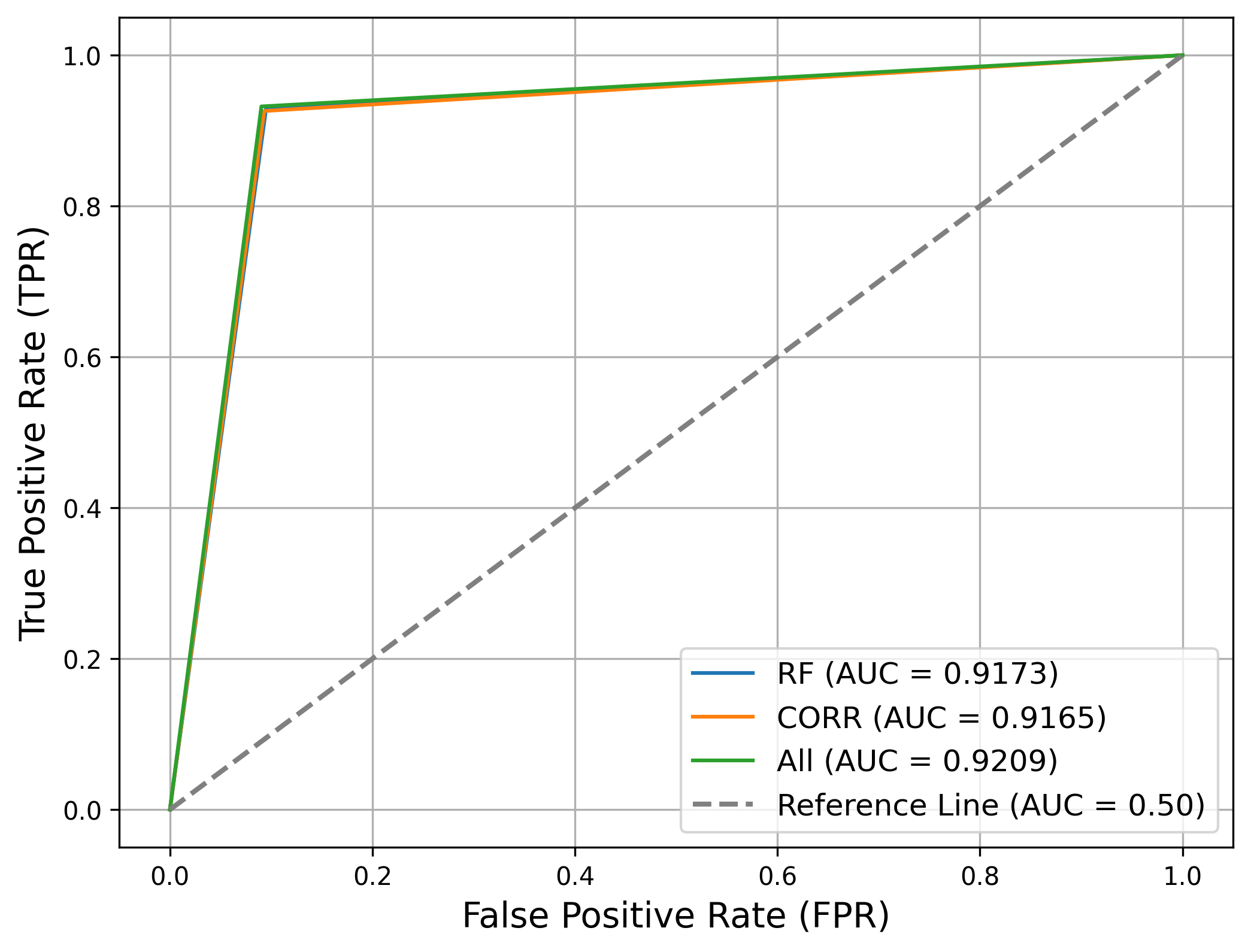}
\caption{ROC-AUC metric for the Random Forest and correlation techniques compared to the one obtained using all features.}
\label{fig:roc_auc}
\end{figure}

To evaluate the implemented models, the validation database previously separated was used to evaluate the implemented models. It is worth mentioning that the validation database contains data that were not used in training and, therefore, are unknown to the models. Figure~\ref{fig:roc_auc} presents the ROC-AUC metric of the RF and correlation techniques compared to the ``All'' approach, using all features. Additionally, Table~\ref{tab:metrics_validation} provides the values of all metrics for better cross-technique analysis.

\vspace{-8pt}
{
\begin{table}[H]%[!h]
\caption{Mean value of assessment metrics of result sets.}
\label{tab:metrics_validation}
\newcolumntype{C}{>{\centering\arraybackslash}X}
\begin{tabularx}{\textwidth}{CCCCCC}
\toprule
\textbf{} & \textbf{Accuracy} & \textbf{Precision} & \textbf{F-1 Score} & \textbf{Recall} & \textbf{ROC AUC}\\\midrule
RF & {0.917} & {0.917} & {0.917} & {0.917} & {0.917} \\
Correlation & {0.916} & {0.916} & {0.916} & {0.916} & {0.917} \\
All & {0.921} & {0.921} & {0.921} & {0.921} & {0.921} \\
\bottomrule
\end{tabularx}
\end{table}
}

\begin{figure}[H]%[h!]
\begin{adjustwidth}{-\extralength}{0cm}
\centering
\includegraphics[width=15.5cm]{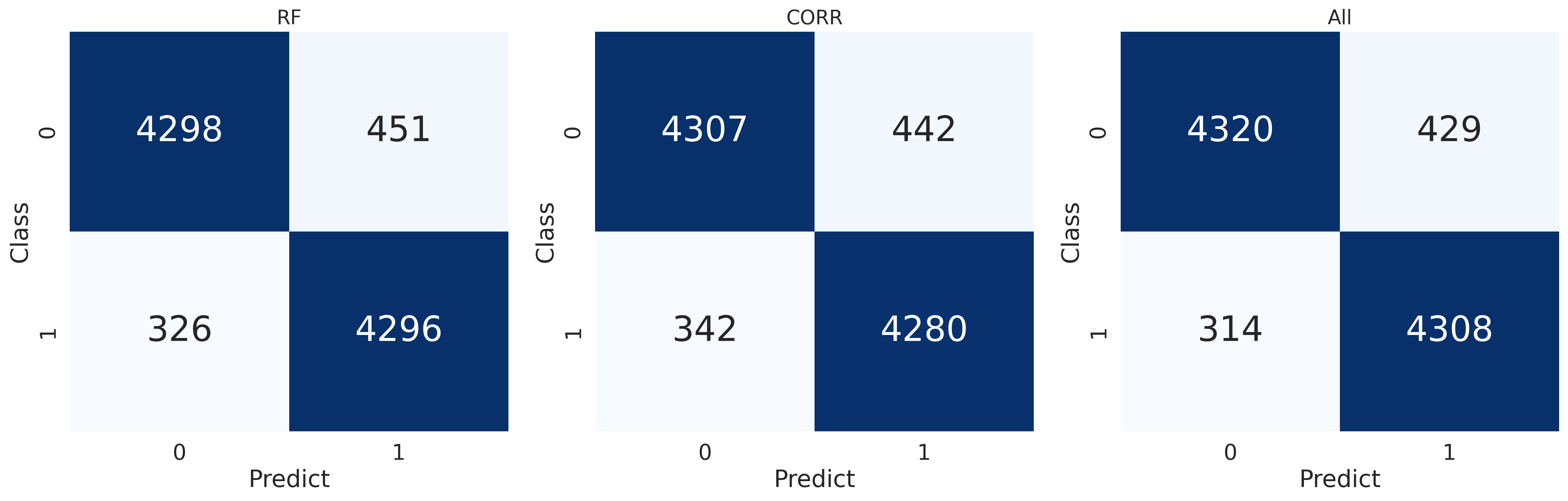}
\end{adjustwidth}
\caption{Confusion matrices for the Random Forest and correlation techniques compared to the one obtained using all features.}
\label{fig:matrices}
\end{figure}

Finally, {Figure}~\ref{fig:matrices} presents the confusion matrix---a confusion matrix is a tabular representation typically used in supervised learning to visualise algorithm performance---%MDPI: Footnote is not permitted in our journal. Please include this paragraph to the maintext.
where the {rows} represent actual labels, and the columns represent predicted labels generated by the analyzed techniques. Classes $0$ and $1$ correspond to background and signal, respectively.

%It can be noted that the use of all resources (All) presents marginally higher values than the other evaluated techniques. However, as seen in Section~\ref{sec:3.1} for model development and training, the RF and Pearson Correlation techniques do not present statistical differences if all characteristics are used. In this sense, given the results presented, it can be stated that using {the set of 75 variables selected by the RF technique}, it is possible to achieve statistically similar results if all 130 characteristics were used, with the advantage of increasing the generalization of the model, reducing the chance of overfitting, decrease training time and increase computational efficiency.

{The comparison of different techniques reveals that the performance metrics obtained for the set of features including all the features available are marginally higher than the ones obtained for the other two methods. However, detailed analysis in \mbox{Section~\ref{sec:3.2}} and \mbox{Section~\ref{sec:CPU_time}} indicates that the set of variables selected by the Random Forest (RF) and correlation techniques do not exhibit statistical differences compared to the set composed by all features while resulting in lower CPU time. %Despite this, these methods differ in their p-values for the Shapiro-Wilk test, favouring the Random Forest for its robustness.
Therefore, employing the set of \mbox{75 variables} selected by the RF technique can achieve statistically similar outcomes to using all 130~characteristics while enhancing model generalisation, reducing overfitting risks, slightly decreasing training time, and boosting computational efficiency.}

%====================================================
%====================================================
%====================================================

\subsection{Discussion on Performance of Random Forest Algorithm}
\label{sec:3.4}

All the ML feature selection algorithms apart from the linear regression perform better than the physics-driven selection presented in Bueno et al.~\cite{Bueno:2023vrg} both in terms of evaluation metrics (see Table~\ref{tab:metrics_geral}) and in terms of statistical compatibility with the complete set of variables (see Table~\ref{tab:t-test}).
{In particular, the values of recall, precision, and F1-score in Table~\ref{tab:metrics_geral} demonstrate improvements in the classification of AMS-02 data compared to the physics-driven selection approach proposed in Bueno et al.~\cite{Bueno:2023vrg}. The higher values scored for recall for the Kbest, Random Forest, and correlation methods with respect to the approach of Bueno et al.~\cite{Bueno:2023vrg} should be interpreted as a more accurate classification of signal events (true positives), while the higher precision values mark a reduction in the false positives. The efficiency in the classification of signal and background events is confirmed in Figure~\ref{fig:roc_auc}, illustrating that a true positive rate (i.e., signal efficiency) of 92\% is achieved with a false positive rate (i.e., background efficiency, denoted as $\epsilon_{bkg}$) of 0.1\%. These results imply a background rejection, defined as $1 - \epsilon_{bkg}$, of 90\%, showcasing the impact of the implemented methods on the data. This performance translates into an efficient separation of the signal and background events, allowing for a cleaner data sample and, hence, a more accurate mass reconstruction and identification of deuteron isotopes.}

Since the Random Forest method is the best-performing selection technique among the ML algorithms compared in this work, the set of features selected by this method will be described in more detail.
As shown in Figure~\ref{piechart:Radar_01} and in Table~\ref{tab:percentagesone}, all features in the ``Charge'' and ``Track Position'' classes were selected by the Random Forest algorithm. A fraction of 80\% of the features was chosen for the ``PMT number'' class, and a similar percentage of 77.8\% was selected from the ``Photoelectrons'' class. The classes ``Beta'' and ``Hit number'' had the lowest feature selection percentages, with 32.5\% and 26.9\%, respectively.

%\begin{figure}[H]%[!h]
%\centering
%\includegraphics[width=1\linewidth,keepaspectratio]{Radar_02.png}
%\caption{Description.}
%\label{piechart:Radar_02}
%\end{figure}

It is worth noting that the classes ``Charge'', ``Photoelectrons'', and ``PMT number'' present minor variability when the results of the different selection methods are compared, confirming that the power of separation of the variables belonging to these classes is independent of the underlying characteristics of a particular selection method and is closely linked to the physical phenomena underlying the RICH detection mechanism.

The features in the classes ``Photoelectrons'' and ``PMT number'' are related to the number of photons detected and used to reconstruct the ring. Because the signal detected in the PMT plane originates from the Cherenkov photons created by the cosmic ray particle that hits the radiator plane, the higher the number of photons, the less likely the signal will be disrupted by noise (Section~\ref{sec:2.1}). Therefore, variables from these two classes are expected to behave differently for background and signal events, resulting in a good classification performance. In particular, events with lower detected ``Photoelectrons'' or ``PMT number'' are more prone to be poorly reconstructed and are expected to be part of the background.

Variables belonging to the ``Charge'' class are indirectly influenced by the same processes. For example, the charge of the reconstructed particle is often higher for background events than for signal events (see Figure~\ref{feat_ex}) due to the inclusion of additional hits in the reconstructed Cherenkov ring and the feature's distribution for background events is more shifted towards high charge values. Likewise, the Kolmogorov probability, which also belongs to the class ''Charge'', shows values for background events on average lower than for signal events since the charge distribution along the ring is not uniform (see Section~\ref{sec:2.1}).

Finally, the percentage of features selected for the ``Track Position'' class has higher variability, but it is still possible to trace  its discrimination power back to the RICH structure. In particular, the impact point of the extrapolated tracker track on the radiator plane is a sensitive observable for this study, as also discussed in Bueno et al.~\cite{Bueno:2023vrg}. Particles impacting some radiator areas (e.g., tile borders) tend to produce fewer detectable Cherenkov photons. They are, therefore, more likely to produce a weaker signal subject to incorrect reconstruction, making it possible to separate background and signal events.

%====================================================
%====================================================
%====================================================

\section{Conclusions}
\label{sec:5}

An efficient reduction in the background consisting of events whose velocities are misreconstructed in the RICH detector is needed to identify positive singly charged cosmic ray isotopes with the AMS-02 detector. ML methods that can be used for this purpose, such as BDTs, are often trained on a set of features selected on the basis of the knowledge of the detector and of the classification task. However, ML algorithms can be used to perform automated feature selection improving the efficiency and accuracy of the analysis.
In this paper, we applied automated feature selection methods to the background reduction analysis for the identification of cosmic ray deuterons with six months of data collected by the AMS-02 detector. {The aim of the study was twofold: to choose among 130 variables associated with the RICH detector the best set of features to reject background events due to misrecontruction of the velocity measured by the RICH detector, and
%to investigate whether ML techniques could enhance the efficiency of the background reduction process compared to traditional physics-driven approach.
to assess the potential of these ML techniques in improving the background reduction efficiency in RICH compared to a traditional physics-driven approach.}
We used five feature selection algorithms widely used in the literature, namely Kbest, Random Forest, linear regression, and correlation, together with the method described in~\cite{Bueno:2023vrg} which has the same scope but uses a set of physically motivated features. We used a boosted decision tree to perform the classification task and a K-fold cross-validation to validate our results.

We assessed the performance of the six methods with different evaluation metrics and found that the methods Kbest, Random Forest, and correlation outperform the approach described in~Bueno et al.~\cite{Bueno:2023vrg} in terms of accuracy, precision, F-1 score, and recall.
Moreover, it is worth noting that only the set of variables selected by the Random Forest and correlation methods do not present statistical differences when compared to the complete set of 130~variables.
Hence, the Random Forest method stands out as the best-performing algorithm, since it shows {similar performance metrics} compared to the complete set of 130~variables while reducing the risk of overfitting and training time and increasing the computational efficiency.
{The results obtained from the ML algorithms demonstrate that there is still some potential for improvement, which is crucial for deuteron identification due to the critical need to reduce the background composed by events with poorly reconstructed RICH velocity, as explained in Section~\ref{sec:1}.
We conclude by investigating whether this ML method also maintains the connection between selected variables and the underlying physical phenomena related to RICH detection mechanisms, finding that the Random Forest algorithm maintains a meaningful link between selected variables and the physics of isotope identification.}
%We conclude by discussing the selection operated by the Random Forest algorithm in the context of the detection and reconstruction techniques used for the RICH detector.

{Finally, the study focuses exclusively on feature selection for BDTs within the context of the AMS-02 experiment. This approach provides a direct and relevant comparison with the previous work of~Bueno et al.~\cite{Bueno:2023vrg}. However, feature selection techniques can also be applied to a broader spectrum of ML models. Therefore, future work will investigate their performance across different ML paradigms, thereby seeking to expand the results~obtained.}

%\end{spacing}

%%%%%%%%%%%%%%%%%%%%%%%%%%%%%%%%%%%%%%%%%%
\vspace{6pt}

%%%%%%%%%%%%%%%%%%%%%%%%%%%%%%%%%%%%%%%%%%
\authorcontributions{Conceptualisation, M.B., M.V. and L.M.; methodology, M.B., L.M. and M.V.; software, L.M.; validation, M.B. and M.V.; formal analysis, M.B. and L.M.; investigation, M.B., L.M. and M.V.; {resources, M.V.}; data curation, M.B. and L.M.; writing---original draft preparation, M.B. and L.M.; writing---review and editing, M.V. and F.B.; visualisation, M.B. and L.M.; supervision, M.V. and F.B.; project administration, M.V.; funding acquisition, M.V. All authors have read and agreed to the published version of the manuscript.} %MDPI: We removed the template content (please turn to the  \href{http://img.mdpi.org/data/contributor-role-instruction.pdf}{CRediT taxonomy} for the term explanation.). Please confirm.

\funding{This work is part of the project ``Cosmic ray antideuterons as a probe for new physics'' with project number OCENW.KLEIN.387 (Budget Number 11680) of the research programme Grant Open Competition Domain Science, which is financed by the Dutch Research Council (NWO). This study was financed in part by the Coordenação de Aperfeiçoamento de Pessoal de Nível Superior---Brasil (CAPES)---Finance Code 001 (Processo 300904/2023-1).}

\dataavailability{The datasets presented in this article are not readily available because property of the AMS Collaboration. Requests to access the datasets should be directed to the ASM Collaboration.} %MDPI: "Not applicable"is only used for review papers or articles for which no new data were created. Normally, the DAS for research articles can state"Data are contained within the article." or "Data are contained within the article and supplementary materials." Please refer to the complete guideline at https://www.mdpi.com/ethics#_bookmark21.

\acknowledgments{\textls[-25]{We would like to thank Eduardo Bueno and Alberto Oliva for insightful discussion.}} %MDPI: Please ensure that all individuals included in this section have consented to the acknowledgement.

\conflictsofinterest{The authors declare no conflicts of interest. The funders had no role in the design of the study; in the collection, analyses, or interpretation of data; in the writing of the manuscript; or in the decision to publish the~results.}
%%%%%%%%%%%%%%%%%%%%%%%%%%%%%%%%%%%%%%%%%%
%% Optional

\appendixtitles{yes} % Leave argument "no" if all appendix headings stay EMPTY (then no dot is printed after "Appendix A"). If the appendix sections contain a heading then change the argument to "yes".
\appendixstart
\appendix
{\section[\appendixname~\thesection]{}} %MDPI: Please confirm if it is unnecessary and can be removed.
%\subsection[\appendixname~\thesubsection]{}
%{The list of all 130 features used in the analysis is reported in the table below. Since some of the variables are stored in arrays, and hence share the array name, we use the notation array\_name[i], indicating the \textit{i-th} element of the array, to denote each feature. E.g. tot\_p[1-5] represents the 5 variables stored in the first 5 elements of the tot\_p array.}

\vspace{-12pt}
\begin{table}[H]%[h!]
\caption{{List of %MDPI: We merged the tables into one. Please confirm.
the features used in the analysis. %Some of the variables are stored in arrays, and hence share the array name. For these, we use the notation array\_name[i], indicating the \textit{i-th} element of the array, to denote each feature. E.g. tot\_p[1-5] represents the 5 variables stored in the first 5 elements of the tot\_p array.
}}
\begin{adjustwidth}{-\extralength}{0cm}
\begin{tabularx}{\fulllength}{p{0.15\linewidth} p{0.75\linewidth}}
\toprule
\textbf{Class}  & \textbf{Feature Description} \\ \midrule
\multirow{12}{*}{Photoelectrons}    & Maximum number of p.e. in a PMT including or excluding crossed PMTs    \\
&  Number of expected p.e. in the absence of a reconstructed Cherenkov ring\\
& Number of p.e. collected inside the Cherenkov ring \\
& Photoelectrons associated to the ring for different windows sizes\\
& Expected number of p.e. for a Z = 1 Cherenkov ring with reconstruction and input parameters of the current event\\
& Expected number of p.e. for a Z = 1 Cherenkov ring with reconstruction and input parameters of the current event and $\beta = 1$ \\
& Number of photons inside the Cherenkov ring \\
& Number of photons outside the Cherenkov ring \\
& Number of p.e. detected in the first 5 PMTs by number of p.e. \\
& Total number of p.e. for $\beta = 1$ hypothesis  \\
& Total number of p.e. out of the ring for a particle with $\beta = 1$ \\ \midrule
\multirow{6}{*}{Charge} & Reconstructed charge (CIEMAT reconstruction)~\cite{Delgado_rich}\\
& Kolmogorov test of the distribution of charge along the ring \\
& Statistical test to check if the hit-by-hit charge is consistent PMT-by-PMT \\
& Expected charge resolution   \\
& Expected charge resolution RMS   \\
& Reconstructed charge (LIP reconstruction)~\cite{barao_rich}\\ \midrule
\multirow{4}{*}{PMT number} & Number of crossed PMTs \\
& Number of PMTs inside the Cherenkov ring\\
& Expected number of PMTs for a Z = 1 Cherenkov ring with reconstruction and input parameters of the current event  \\

\midrule
%		\end{tabularx}
%	\end{adjustwidth}
%	\noindent
%\end{table}
%
%\begin{table}[H]%[h!]
%%\caption{Number (and percentages) of the feature selected for each class by the different methods}
%\begin{adjustwidth}{-\extralength}{0cm}
%        \begin{tabularx}{\fulllength}{p{0.15\linewidth} p{0.75\linewidth}}
%			%\toprule
%            \textbf{Class}  & \textbf{Feature Description} \\ \midrule
\multirow{8}{*}{Hit Number}     & Total number of hits\\
& Number of hits in the first 5 PMTs by number of p.e. \\
& Total number of hits compatible with $\beta = 1$ hypothesis (direct and reflected) \\
& Number of hits out of the Cherenkov ring compatible with $\beta = 1$ hypothesis (direct and reflected) \\
& Number of hits inside the Cherenkov ring    \\
& Number of hits which are consistent with reflected photons \\
& Distribution for unused hits which do not belong to the PMTs crossed by a charged particle \\
& Number of hits outside the Cherenkov ring  \\

\midrule
\multirow{7}{*}{Track Position} &  Distance from the tracker track [cm] for the first 5 PMTs by number of p.e. \\
&  PMT number for the first 5 PMTs by number of p.e.  \\
&  Tile ID for the tile crossed by the particle  \\
&  Impact point of the tracker track to the radiator entrance [cm]  \\
&  Theta of the tracker track to the radiator entrance [rad]  \\
&  Phi of the tracker track to the radiator entrance [rad] \\
&  Distance of the tracker track impact point on the radiator to the border of the radiator tile  \\
\bottomrule
\end{tabularx}
\end{adjustwidth}
\end{table}
\begin{table}[H]\ContinuedFloat

\caption{\textit{Cont.}}
\begin{adjustwidth}{-\extralength}{0cm}%\centering
\begin{tabularx}{\fulllength}{p{0.15\linewidth} p{0.75\linewidth}}
\toprule
\textbf{Class}  & \textbf{Feature Description}   \\
\midrule
\multirow{12}{*}{Beta}  &  Reconstructed particle velocity $\beta$ (CIEMAT reconstruction)~\cite{Delgado_rich}\\
&  Expected velocity resolution \\
&  Expected resolution RMS \\
&  Raw $\beta$\\
&  Refit $\beta$\\
&  $\beta$ corrected for impact point and direction\\
&  Clusters of signal used for the reconstruction of the (particle) velocity $\beta$.\\
&  Size of first 10 clusters\\
&  Average beta of first 10 clusters  \\
&  RMS of first 10 clusters \\
&  Reconstructed particle velocity $\beta$ (LIP reconstruction)~\cite{barao_rich}\\
&  $\beta$ estimated from rigidity   \\
\bottomrule
\end{tabularx}
\end{adjustwidth}
\noindent
\label{tab:features_list}
\end{table}
\unskip
\begin{table}[H]%[h!]
\caption{{Parameters used in each feature selection technique used in this work, together with the corresponding Python library/method  used to implement them.}}
\begin{adjustwidth}{-\extralength}{0cm}
%\newcolumntype{C}{>{\centering\arraybackslash}X}
\newcolumntype{L}[1]{>{\hsize=#1\hsize\RaggedRight} X}
\begin{tabularx}{\fulllength}{L{0.15} L{0.40} L{0.45} }
%\begin{tabularx}{\fulllength}{p{0.10\linewidth} p{0.45\linewidth} p{0.30\linewidth}}
\toprule
\textbf{Techniques}  & \textbf{Parameters}  & \textbf{Library/Method} \\ \midrule
Kbest       &  score\_fun=f\_classif, k=10, X=sparse matrix of shape (n\_samples, n\_features), y=array-like of shape (n\_samples,) %MDPI: Please confirm if the space is necessary and can be added. The following highlights are the same.
& sklearn.feature\_selection.SelectKBest \\ \hline
RF          &  n\_estimators=100, criterion=gini, max\_depth=None, min\_samples\_split=2, min\_samples\_leaf=1, min\_weigh\_fraction\_leaf=0.0, max\_features=sqrt, max\_leaf\_nodes=None, min\_impurity\_decrease=0.0, bootstrap=True, oob\_score=False, n\_jobs=None, random\_state=None, verbose=0, warm\_start=False, class\_weight=None, ccp\_alpha=0.0, max\_samples=None, monotonic\_cst=None & sklearn.ensemble.RandomForestClassifier \\ \hline
Linear      &  estimator=LinearRegression(), fit\_intercept=True, copy\_X=True, positive=False, step=1, min\_features\_to\_select=1, cv=10, scoring=None, verbose=0, n\_jobs=None, importance\_getter=auto & sklearn.feature\_selection.RFECV \\ \hline
Correlation & method=pearson, min\_periods=None, numeric\_only=False, threshold=0.005 & pandas.DataFrame.corr \\
\bottomrule
\end{tabularx}
\end{adjustwidth}
\noindent
\label{tab:parameters}
\end{table}
\vspace{-12pt}
%\newpage
%{\section[\appendixname~\thesection]{Metrics uncertainties}}
%{We report below the uncertainties on the performance metrics for the sets of selected features.}

\begin{table}[H]%[!h]
\caption{{Uncertainties on assessment metrics for the selected features.}}
%\begin{center}
\label{tab:uncertainties}
%\begin{tabular}{ccccccc}\hline
\begin{adjustwidth}{-\extralength}{0cm}
\newcolumntype{C}{>{\centering\arraybackslash}X}
\begin{tabularx}{\fulllength}{CCCCCC}
\toprule
\textbf{} & \textbf{Accuracy} & \textbf{Precision} & \textbf{F-1 Score} & \textbf{Recall} & \textbf{ROC AUC} \\\midrule
Kbest & 0.004137 & 0.005627 & 0.004245 & 0.008718 & 0.003883 \\
RF & 0.006519 & 0.005536 & 0.006671 & 0.010665 & 0.003592 \\
Linear & 0.008393 & 0.005494 & 0.006611 & 0.010761 & 0.008763\\
Correlation & 0.006291 & 0.007005 & 0.006229 & 0.008382 & 0.003002\\
Bueno et al. & 0.005628 & 0.007487 & 0.005734 & 0.010649 & 0.005049 \\
All & 0.006056 & 0.005169 & 0.006127 & 0.008921 & 0.003326 \\\bottomrule
%\end{tabular}
%\end{center}
\end{tabularx}
\end{adjustwidth}
\end{table}

%\section[\appendixname~\thesection]{}
%All appendix sections must be cited in the main text. In the appendices, Figures, Tables, etc. should be labeled, starting with ``A''---e.g., Figure A1, Figure A2, etc.

%%%%%%%%%%%%%%%%%%%%%%%%%%%%%%%%%%%%%%%%%%

\begin{adjustwidth}{-\extralength}{0cm}
%\printendnotes[custom] % Un-comment to print a list of endnotes

\reftitle{References}

% Please provide either the correct journal abbreviation (e.g., according to the “List of Title Word Abbreviations” http://www.issn.org/services/online-services/access-to-the-ltwa/) or the full name of the journal.
% Citations and References in Supplementary files are permitted provided that they also appear in the reference list here.
% For the MDPI journals use author-date citation, please follow the formatting guidelines on http://www.mdpi.com/authors/references
% To cite two works by the same author: \citeauthor{ref-journal-1a} (\citeyear{ref-journal-1a}, \citeyear{ref-journal-1b}). This produces: Whittaker (1967, 1975)
% To cite two works by the same author with specific pages: \citeauthor{ref-journal-3a} (\citeyear{ref-journal-3a}, p. 328; \citeyear{ref-journal-3b}, p.475). This produces: Wong (1999, p. 328; 2000, p. 475)

%\bibliographystyle{ieeetr}

\PublishersNote{}
\end{adjustwidth}
\end{document}